\def \yskip{\penalty-50\vskip3pt plus 3pt minus 2pt}
\def \reference{\par \yskip \noindent \hangindent .4in \hangafter 1}
\def \abc#1#2#3#4 {\reference#1, {\sl#2}, {\bf#3}, #4}
\def \blank {\lower 5pt\hbox to 0.75in{\hrulefill}}

\def \AU{~\rm{AU}}
\def \yrs{~\rm{yrs}}
\def \yr{~\rm{yr}}

\documentstyle[11pt,aasms,tighten,flushrt]{article}

\begin{document}
\small

\setcounter{page}{1}
\begin{center} \bf 
ECCENTRIC ORBITS OF CLOSE COMPANIONS TO
ASYMPTOTIC GIANT BRANCH STARS
\end{center}

\begin{center}
Noam Soker\\
Department of Physics, University of Haifa at Oranim\\
Oranim, Tivon 36006, ISRAEL \\
soker@physics.technion.ac.il 
\end{center}


\begin{center}
\bf ABSTRACT
\end{center}

 I propose that the relatively high eccentricity
$0.1 \lesssim e \lesssim 0.4$ found in some tidally strongly interacting
binary systems, where the mass-losing star is an evolved giant star,
e.g., an asymptotic giant branch star, is caused by an enhanced mass
loss rate during periastron passages.
 Tidal interaction by itself will circularize the orbits of these
systems in a relatively short time, hence a mechanism which increases
the eccentricity on a shorter time scale is required.
 The proposed scenario predicts that the nebula formed by the mass
loss process possesses a prominent departure from axisymmetrical
structure.
 Such a departure is observed in the Red Rectangle, which
has a central binary system, HD 44179, with an orbital period of
$T_{\rm orb} = 318$ days, and an eccentricity of $e=0.38$.
 



\section{INTRODUCTION}
 
 In recent years it has been found that close binary companions to
asymptotic giant branch (AGB) stars can influence the 
chemistry of the circumbinary shells (or disks) and of the post-AGB stars 
(e.g., Van Winckel, Waelkens, \& Waters 1995), and the eccentricities 
of the orbits (e.g., Waelkens {\it et al.} 1996). 
(In this paper I refer by ``close binary systems'' to  
systems with orbital periods of $T_{\rm orb} \sim 1 \yr$, i.e., 
there is a very strong tidal interaction between the AGB star and its 
companion. By wide systems I refer to systems where the tidal 
interaction is very weak.)  
 Van Winckel {\it et al.} (1998) suggest that all post-AGB star with
peculiar abundances are in close binaries.
 The peculiar abundances, i.e., depletion of Fe peak elements and some other
elements (e.g.,  Waters, Trams, \& Waelkens 1992;
Van Winckel {\it et al.} 1999), is explained by accretion back from the
circumbinary disk of gas depleted of dust (Waters {\it et al.} 1992), 
since the dust is expelled more efficiently from the disk by radiation pressure.
 The disk helps by providing a slow accretion rate and a low density to 
prevent an efficient drag between the gas and dust particles 
(Waters {\it et al.} 1992).
 It is not clear, though, that a Keplerian circumbinary disk is a necessary
condition for the separation of gas from dust.
In an earlier paper (Soker 2000) I argued that a dense slowly expanding
equatorial flow may have the same effect, while having the advantage
that it does not require a huge amount of angular momentum as
a Keplerian disk does.
 
 Another peculiar property found among many of these binary systems
is a high eccentricity (e.g., Van Winckel 1999, and references therein),
despite the very strong tidal interaction between the AGB stars and
their companions.
Examples are:
 {\bf HD 44179}, which is located at the center of the
Red Rectangle, a bipolar proto planetary nebula (PN), has an 
orbital period of $T_{\rm orb} = 318$ days, a semimajor axis of 
$a \sin i = 0.32 \AU$, and an eccentricity 
of $e=0.38$ (Waelkens {\it et al.} 1996; Waters {\it et al.} 1998). 
 {\bf AC Her}, with $T_{\rm orb} = 1194$ days, $a \sin i = 1.39 \AU$, 
and $e=0.12$ (Van Winckel {\it et al.} 1998). 
 {\bf 89 Herculis}, with $T_{\rm orb} = 288$ days, 
and $e=0.19$ (Waters {\it et al.} 1993). 
 {\bf HR 4049}, with $T_{\rm orb} = 429$ days, 
$a \sin i = 0.583 \AU$, and $e=0.31$ (Van Winckel {\it et al.} 1995).

 Van Winckel {\it et al.} (1995) mention in one sentence in their discussion
that ``. . . the large eccentricities suggest that mass loss may have  
started at periastron only, thus increasing the eccentricity still.''
 However, they did not carry out any quantitative study, but abandoned
this explanation in favor of the ``external disk'' mechanism
(Waelkens {\it et al.} 1996; Waters {\it et al.} 1998). 
 What I term ``external disk'' mechanism is the tidal interaction between
the binary system and the circumbinary disk, which is the model to
explain eccentricities in young stellar binaries 
(Artymowicz {\it et al.} 1991; Artymowicz \& Lubow 1994).
 
 In the present paper I claim that higher mass loss rate during
periastron passage can indeed explain the high eccentricities observed
in the systems mentioned above. 
 I am motivated by earlier results that variation in the mass loss 
rate and the mass transfer rate with orbital phases in eccentric orbits 
can explain the formation of multiple shells in PNs
(Harpaz, Rappaport, \& Soker 1997; in that paper, though, the orbital 
periods are $\sim 100 \yrs$ rather than $1 \yr$), and the displacement 
of the the central stars from the centers of PNs 
(Soker, Rappaport, \& Harpaz 1998).
 In the next section I describe the proposed model and the
time scales involved, while in $\S 3$ the external disk mechanism
proposed in other studies (e.g., Waelkens {\it et al.} 1996) is discussed.
A short summary is in $\S 4$. 

\section{ECCENTRICITY EVOLUTION DUE TO MASS LOSS} 
  
 The eccentricity $e$ is reduced by tidal forces on a time scale called
the circularization time and is defined as 
$\tau_{\rm circ} \equiv -e/\dot e$.
 In the common tidal model in use, the equilibrium tide mechanism
(Zahn 1977; 1989), the circularization time is given by 
(Verbunt \& Phinney 1995)
\begin{eqnarray}
\tau_{\rm {circ}} =  
1.2 \times 10^4
{{1}\over{f}} 
\left( {{L} \over {2000 L_\odot}} \right)^ {-1/3}
\left( {{R} \over {200 R_\odot}} \right)^ {2/3}
\left( {{M_{\rm {env}}} \over {0.5M_1}} \right)^ {-1} 
\left( {{M_{\rm {env}}} \over {0.5M_\odot}} \right)^ {1/3} \nonumber \\
\times \left( {{M_2} \over {M_1}} \right)^ {-1}
\left( 1+ {{M_2} \over {M_1}} \right)^ {-1}
\left( {{a} \over {3R}} \right)^ {8}
\yr ,
\end{eqnarray}
where $L$, $R$ and $M_1$ are the luminosity, radius, and total mass of the 
primary AGB star, $M_{\rm {env}}$ is the primary's envelope mass, 
and $f \simeq 1$ is a dimensionless parameter. 
 The synchronization time is related to the circularization time
by the expression 
$\tau_{\rm {syn}} \simeq  (1+M_2/M_1)(M_2/M_1)^{-1}(I/M_1R^2) 
(R/a)^2 \tau_{\rm {cir}} $, where $I$ is the primary's moment of inertia.
Approximating the envelope density profile of stars on the upper  
RGB and AGB by $\rho \propto r^{-2}$, where $r$ is the radial distance
from the star's center, I find $I=(2/9) M_{\rm env} R^2$. 
 Substituting this in the expression for the synchronization time I find
\begin{eqnarray}
\tau_{\rm {syn}} =  
150 
{{1}\over{f}} 
\left( {{L} \over {2000 L_\odot}} \right)^ {-1/3}
\left( {{R} \over {200 R_\odot}} \right)^ {2/3}
\left( {{M_{\rm {env}}} \over {0.5M_\odot}} \right)^ {1/3}
\left( {{M_2} \over {M_1}} \right)^ {-2}
\left( {{a} \over {3R}} \right)^ {6}
\yr.
\end{eqnarray}
 The short synchronization time means that the AGB star will spin with an
angular velocity of $\omega \simeq (R/a)^{3/2} \omega_{\rm Kep}$,
where $\omega_{\rm Kep}$ is the Keplerian velocity on the 
equatorial line of the primary. 
 The high angular velocity and close companion will lead to an enhanced
mass loss rate in the equatorial plane (Mastrodemos \& Morris 1999).

 The change in eccentricity due to an isotropic mass loss 
(the derivation is applicable for an axisymmetric mass loss as well; 
mass transfer will be discussed later) is given by 
(Eggleton 2000) 
\begin{eqnarray}
\delta e = 
{{\vert \delta M \vert} \over {M}} (e + \cos \theta), 
\end{eqnarray}
where $\delta M$ is the mass lost from the binary in the stellar wind 
at the orbital phase $\theta$ (hence $\delta M < 0$), $M$ is the total
mass of the binary system, and $\theta$ is the polar angle, measured
from periastron, of the position vector from the center of mass to
the secondary.
 The derivation of equation (3) assumes that
$\delta M(\theta)=\delta M (-\theta)$. 
 To derive a time scale, I assume that in addition to its constant mass
loss rate over the orbital motion $\dot M_w$, the primary AGB star
loses an extra mass $\delta M_p$ in a short time during the periastron
passage, $\cos \theta =1$.
 The total mass being lost in one orbital period 
$T_{\rm orb}$, is $\Delta M_o = \dot M_w T_{\rm orb} + \delta M_p$. 
I define the fraction of the mass being lost at periastron
\begin{eqnarray}
\beta \equiv {{\delta M_p}\over{\Delta M_o}}.
\end{eqnarray}
  The constant mass loss rate $\dot M_w$ does not change the eccentricity.
 Hence, the change in the eccentricity in one orbital period is 
$\delta e = (1+e) (\vert \delta M_p \vert / M )$.
 Over a time much longer than the orbital period we can write
for the rate of change of the eccentricity
\begin{eqnarray}
{{de}\over{dt}} = - (1+e) \beta {{\dot M}\over{M}} 
\end{eqnarray}
 where the total mass loss rate is
$\dot M = \Delta M_o/T_{\rm orb} = \dot M_w + \delta M_p/T_{\rm orb}$.
  I define the time scale for change in eccentricity due to enhanced
periastron mass loss as 
\begin{eqnarray}
\tau_p \equiv {{e}\over{de/dt}} 
= 4 \times 10^3 \beta^{-1}
\left[ {{e}\over{0.2(1+e)}} \right]
\left( {{M_1+M_2} \over {2M_\odot}} \right)  
\left( {{\vert \dot M_o \vert} \over {10^{-4} M_\odot \yr^{-1}}} \right)^{-1}  
\yr.
\end{eqnarray}

Under the assumption that the fraction of mass lost at 
periastron passage $\beta$ does not change during the 
evolution, we can integrate equation (5) to yield
\begin{eqnarray}
{{1+e}\over{1+e_i}} = 
\left( {{M_i}\over{M}} \right)^{\beta}, 
\end{eqnarray}
where $e_i$ and $M_i$ are the initial eccentricity and total 
mass, respectively. 
 Both the initial over final mass and $\beta$ can vary continuously
among different systems.
 It is useful, however, to examine two extreme cases.
\newline
{\bf Systems with $\beta \simeq 1$:}
When most of the mass is being lost during a periastron passage, then
$\beta \simeq 1$.
 Such can be the case for low mass AGB stars and/or stars not yet on
the upper AGB, so that the mass loss rate is low, and the secondary,
via direct gravitational effects, increases substantially the mass
loss rate during periastron passages.
 For systems with $\beta \simeq 1$, a moderate amount of total mass loss
can substantially increase the eccentricity.
 For example, for initial (initial means at the beginning
of the interaction, not on the main sequence) masses of
$M_{1i}=1 M_\odot$ and $M_2 = 0.6 M_\odot$, and $e_i \ll 1$, if the
primary loses its entire envelope of $0.4 M_\odot$, the eccentricity
will be given by $1+e \simeq 1.6/1.2 = 1.33$ or $e \simeq 0.3$.
 For $\beta =0.5$ we get $e \simeq 0.15$ for the same mass loss evolution. 
The circumbinary wind (or nebula) is expected to have two 
prominent properties.
First, since most of the mass, at least in the equatorial plane,
is lost via a dynamical interaction between
the two binary stars, and not by radiation pressure on dust,
the wind expansion velocity in the equatorial plane will be very 
low (Soker 2000). 
Second, since most of the mass, or a substantial fraction of it, is lost
during a periastron passage when the mass-losing star always moves in
the same direction, the nebula is expected to possess a large
degree of departure from axisymmetry (Soker {\it et al.} 1998). 
 Such a process may explain the departure from axisymmetry observed
in the Red Rectangle.
 The Red Rectangle has a close eccentric binary system,
with $T_{\rm orb} = 318$ days and $e=0.38$.
The general structure of the Red Rectangle is highly axisymmetrical, up
to $\sim 1 ^\prime$ from the central star (e.g., Van Winckel 2000).
However, the 10 $\mu$m map presented by Waters {\it et al.}
(1998; their fig. 3) shows a clear departure from axisymmetry at scales
of $\sim 5 ^{\prime \prime}$ from the central star. 
 Their contour map shows that the equatorial matter is more
extended in the west side.
\newline
{\bf Systems with $\beta \ll 1$:}
 In these cases the total mass lost by the system should be  
larger, or only slightly smaller, than the total binary final
mass in order for the eccentricity to be $e \gtrsim 0.1$. 
 This is expected for systems where the mass-losing star reaches the
upper AGB, hence has a strong wind.
 The rotation, because of synchronization with the orbital motion (eq. 2),
can further increase the mass loss rate.
 Contrary to the previous case, the wind will have an expansion
velocity typical for AGB stars, and the departure from axisymmetry 
will be small.
 Only in the equatorial plane might the flow be slower, since the
mass lost at periastron via dynamical interaction may not be
accelerated to high velocities (Soker 2000). 
 The departure from axisymmetry might still be large enough to be
detected by observations. 

 The tidal circularization time  (eq. 1) and the time scale for
the increase of eccentricity due to periastron mass loss
(eq. 6) have a different dependence on the eccentricity,
as well as on other parameters.
 We can find the conditions for the eccentricity to grow by requiring
that the time scale given by equation (6) be shorter than the
circularization time given by equation (1).
 For this purpose, we can neglect the dependence on the luminosity,
radius, and envelope mass in equation (1).
  Neglecting the dependence on the envelope mass is justified since
from equation (7) it emerges that for the eccentricity to change
by $e \gtrsim 0.1$, the envelope mass to be lost should be
$\gtrsim 0.2 M_\odot$. 
 Using these approximations and $e \ll 1$, and taking
$M_1=M_2 =1 M_\odot$, we find the condition for the eccentricity to grow
\begin{eqnarray}
\left( {{a} \over {3R}} \right)^ {8}
\gtrsim
\beta^{-1}
\left({{e}\over{0.2}} \right)
\left( {{\vert \dot M_o \vert} \over {10^{-4} M_\odot \yr^{-1}}} \right)^{-1}.  
\end{eqnarray}
 In general, it is expected that $a \gtrsim 3 R$.
This is since for $e\simeq 0.3$ the periastron distance for $a=3 R$
is $a_p \simeq 2 R$, and a Roche lobe over flow is likely to occur
at these distances.
 The mass loss process increases the orbital separation, increasing further
the ratio $a/R$.
If the eccentricity does not grow much, and the continuous mass loss
rate $\dot M_w$ increases as the AGB star's envelope mass decreases,
then the system changes from a $\beta \simeq 1$ system into
a $\beta \ll 1$ system.

 Until now I have considered only mass loss, neglecting mass transfer. 
When the primary's radius to orbital separation ratio increases  
(as the primary expands along the AGB) to $(R/a) \gtrsim 0.5$, 
mass transfer, e.g., due to Roche lobe overflow, becomes important.  
 The change in eccentricity due to a mass  $\delta M_{\rm tran}$
transferred from the primary to the secondary is given by (Eggleton 2000)
\begin{eqnarray}
\delta e = 2 \delta M_{\rm tran} 
\left( {{1}\over{M_1}} - {{1}\over{M_2}} \right) (e+\cos \theta).
\end{eqnarray}
 Since enhanced mass transfer is expected during periastron passage,
we see that the eccentricity will increase if $M_1 < M_2$,
as is required for stable mass transfer and is found for these systems
(e.g., HD 44179 in the Red Rectangle; Waelkens {\it et al.} 1996). 
 Because of the extended envelope of AGB stars, the mass transferred 
can be several times the mass lost in one orbital period. 
 Therefore, the change in eccentricity due to mass transfer 
may become important when the mass-losing star becomes less massive than
the accretor. 
 The tidal interaction, though, will become stronger as well at
these small orbital separations.

 A final comment to this section is that because of their close companions,
these post-AGB stars did not evolve on a canonical AGB track
(Van Winckel, H., private communication).
However, it is still expected that they had a high mass loss rate
in their recent past.

\section{COMPARISON WITH THE EXTERNAL DISK PROCESS} 

 Waelkens {\it et al.} (1996) attributed the eccentricity of post-AGB
close binaries to the external disk mechanism.
 In the external disk mechanism, which was developed for young binary
systems (e.g., Artymowicz \& Lubow 1994), there is a resonance
interaction between the binary system and an external disk,
mainly with the disk material closer than $\sim 6 a$ to the 
binary system (e.g.,  Artymowicz {\it et al.} 1991). 
 The eccentricity increases at a rate given by 
(e.g.,  Artymowicz {\it et al.} 1991) 
\begin{eqnarray}
\dot e \simeq 1.9 \times 10^{-3} 
\left( {{M_{\rm disk}}\over{M}}\right)
\left( {{2 \pi}\over{T_{\rm orb}}} \right),
\end{eqnarray}
where $M_{\rm disk}$ is the mass of the disk inner to $\sim 6 a$.

 The large uncertainty for post-AGB stars, or any post-main sequence
mass-losing star, is the disk mass within $\sim 6 a$.
 There is strong evidence for molecular disks in several
AGB or post-AGB stars, but these disks have typical radii of
several$\times 100 \AU$ (Jura \& Kahane 1999).
 The inner boundaries of the disks can be much closer to the binary system,
typically $\sim 20 \AU$, or even $5-10 \AU$ in some cases (Van Winckel,
private communication).
 In the Red Rectangle the inner boundary is modeled to be at
$\sim 15 \AU$ (Waelkens {\it et al.} 1996), which
is more than six times the semi-major axis of its central binary system
HD 44179.
 In 89 Herculis the inner boundary of the circumbinary material,
whether a disk or not, is at $r_i \sim 40 \AU$
(Waters {\it et al.} 1993).
 Waters {\it et al.} (1993) also estimated the disk mass to
be $6 \times 10^{-4} M_\odot$.
 The circumbinary material extends to at least several tens of AU
(Alcolea \& Bujarrabal 1991), therefore the mass inside a radius of $6 a$
will in any case be very small, $< 10^{-4} M_\odot$, so that
the time scale for eccentricity increase will be very long.

 The conclusion from this section seems to be that the disk's mass
close to the binary systems, i.e., within $\sim 6 a$, is
too small in these systems to account for the high eccentricity
via the external disk mechanism. 
 Since these systems are post-AGB stars (or similar objects),
it is still possible that during their AGB phase the disk mass was
high enough for the external disk mechanism to be efficient.
 This possibility requires further examination. 

\section{SUMMARY} 

 In the present paper I propose that the relatively high eccentricity
$e \lesssim 0.4$, of tidally strongly interacting binary systems
where the mass-losing star is an AGB star, a post-AGB star,
or a similar object, is caused by an enhanced mass loss rate 
during periastron passages.
 Tidal interaction by itself will circularize the orbits of these
systems in a relatively short time.
 Therefore, a mechanism which increases the eccentricity on a shorter
time scale is required.
 Waelkens {\it et al.} (1996) suggested that the mechanism is an
interaction with a circumbinary disk.
 The interaction occurs mainly with the inner regions of the disk,
within $\sim 6$ times the binary orbital separation
(e.g., Artymowicz {\it et al.} 1991). 
 In $\S 3$ above I argued that such disks, if they exist, do not
contain enough mass in their inner regions to explain the high
eccentricity.
 Instead, I showed in $\S 2$ that an increase in the mass loss rate
at periastron passages can explain these eccentricities.

 Both the interaction with a circumbinary disk mechanism 
and the enhanced periastron mass loss rate mechanism predict much 
higher mass loss rate in the equatorial plane, via the strong binary
interaction (Mastrodemos \& Morris 1999), leading to the
formation of a bipolar nebula, e.g., the Red Rectangle. 
 But each has another strong prediction.
 The external disk mechanism predicts the presence of a relatively
massive disk close to the binary system.
  So far there is evidence only for more extended disks.
It remains to be shown, theoretically and observationally,
that such disks exist (and the material is indeed in a disk and 
not in an outflow).
 The model proposed in the present paper predicts a prominent departure
from axisymmetry, since the mass-losing star moves in the same direction
at each enhanced mass loss phase during the periastron passage 
(Soker {\it et al.} 1998). 
  The Red Rectangle possesses a clear departure from axisymmetry.
  This case was discussed in $\S 2$.

 Another interesting object is the binary system AFGL 4106, which has a 
nebula with a clear departure from axisymmetry 
(Van Loon {\it et al.} 1999; Molster {\it et al.} 1999). 
 However, this binary system is not a post-AGB system, but
is composed of two massive stars (Molster {\it et al.} 1999),
with an orbital period of less than 4500 days.
Its eccentricity has not been determined yet, and according to
the model presented here, it is very likely that the eccentricity 
of the binary system AFGL 4106 is $e>0.1$.


\bigskip

{\bf ACKNOWLEDGMENTS:} 
I thank Hans Van Winckel for many helpful discussions and comments.
 This research was supported in part by a grant from the University
of Haifa and a grant from the Israel Science Foundation. 


\end{document}